\documentclass{aa}  
\usepackage{color}
\usepackage{comment}
\usepackage{natbib}
\usepackage{url}
\usepackage{hyperref}
\usepackage{subfigure}
\usepackage{xspace}
\usepackage[percent]{overpic}
\usepackage{subfloat}
\usepackage{lineno}
\usepackage{amsmath,amstext}
\usepackage{balance}
\usepackage{booktabs}
\usepackage{multirow}
\usepackage{hyperref}
\usepackage[normalem]{ulem}

\newcommand{\ixpe}{{\it IXPE}\xspace}
\newcommand{\fermi}{{\it Fermi}\xspace}

\newcommand{\sifap}{{\it SiFAP2}\xspace}

\graphicspath{{./}{figures/}}

\begin{document}

   \title{Searching for optical pulsations from spider millisecond pulsars with the fast photometer \sifap}
    \titlerunning{Optical pulsations from millisecond pulsars}
\authorrunning{Malacaria et al.}

\author{
C.~Malacaria\inst{\ref{in:INAF-OAR}}\fnmsep\thanks{\email{christian.malacaria@inaf.it}}
\and A.~Papitto\inst{\ref{in:INAF-OAR}} 
\and F.~Ambrosino\inst{\ref{in:INAF-OAR}}
\and C. Ballocco\inst{\ref{in:INAF-OAR},\ref{in:Sapienza}}
\and R. La Placa\inst{\ref{in:INAF-OAR}}
\and G. Illiano\inst{\ref{in:ICE}, \ref{in:IEEC}, \ref{in:INAF-MI}}
\and A. Miraval Zanon\inst{\ref{in:ASI}}
\and L.~Stella\inst{\ref{in:INAF-OAR}}
\and M.~Turchetta\inst{\ref{in:Norway}}
\and J.~D.~Turner\inst{\ref{in:INAF-OAR}}
\and C.~Blanchard \inst{\ref{in:CNRS}}
\and I.~Cognard \inst{\ref{in:CNRS}}
\and L.~Guillemot \inst{\ref{in:CNRS},\ref{in:Nançay}}
\and M.~Burgay \inst{\ref{in:INAF-CA}}
\and A.~Burtovoi \inst{\ref{in:UniFlo},\ref{in:INAF-TO}}
\and A. Sanna \inst{\ref{in:UniCA}}
\and S. Campana \inst{\ref{in:INAF-MI}}
\and M. Cecconi \inst{\ref{in:Fundación}}
\and A. Ghedina \inst{\ref{in:Fundación}}
\and A. Possenti \inst{\ref{in:INAF-CA}}
\and D.~Torres \inst{\ref{in:ICE},\ref{in:IEEC},\ref{in:ICREA}}
}
\institute{
INAF Osservatorio Astronomico di Roma, via di Frascati 33, I–00078, Monteporzio Catone, Roma, Italy \label{in:INAF-OAR} 
\and Dipartimento di Fisica, Sapienza Università di Roma, Piazzale Aldo Moro 5, I-00185 Rome, Italy \label{in:Sapienza}
\and Institute of Space Sciences (ICE, CSIC), Campus UAB, Carrer de Can Magrans s/n, 08193 Barcelona, Spain \label{in:ICE}
\and Institut d’Estudis Espacials de Catalunya (IEEC), 08034 Barcelona, Spain \label{in:IEEC}
\and INAF - Osservatorio Astronomico di Brera, Via Bianchi 46, I-23807, Merate (LC), Italy \label{in:INAF-MI}
\and ASI - Agenzia Spaziale Italiana, Via del Politecnico snc, I-00133 Rome, Italy \label{in:ASI}
\and Department of Physics, Norwegian University of Science and Technology, NO-7491 Trondheim, Norway \label{in:Norway}
\and LPC2E, OSUC, Univ Orleans, CNRS, CNES, Observatoire de Paris, F-45071 Orleans, France \label{in:CNRS}
\and Observatoire Radioastronomique de Nançay, Observatoire de Paris, Université PSL, Université d’Orléans, CNRS, 18330 Nançay, France \label{in:Nançay}
\and INAF - Osservatorio Astronomico di Cagliari, Via della Scienza 5, I-09047 Selargius (CA), Italy \label{in:INAF-CA}
\and University of Florence - Physics and Astronomy Department, Via Sansone 1, I-50019, Sesto Fiorentino (FI), Italy \label{in:UniFlo}
\and INAF - Turin Astrophysical Observatory, Via Osservatorio 20, I-10025 Pino Torinese (TO), Italy \label{in:INAF-TO}
\and Dipartimento di Fisica, Università degli Studi di Cagliari, SP Monserrato-Sestu km 0.7, I-09042 Monserrato, Italy \label{in:UniCA}
\and Fundación Galileo Galilei - INAF, Rambla Josè Ana Fernández Pèrez, 7, E–38712 Breña Baja, TF - Spain \label{in:Fundación}
\and Institució Catalana de Recerca i Estudis Avançats (ICREA), 08010 Barcelona, Spain \label{in:ICREA}
}

   \date{\today}

 
  \abstract
   {Optical pulsations from millisecond pulsars (MSPs) have recently been detected in a transitional MSP and an accreting MSP. Evidence for optical pulsations has also been found for a rotation-powered redback MSP. Observations suggest that optical pulsed emission is produced by different mechanisms depending on whether an accretion disk is present in the system.}
   {We explored a sample of rotation-powered MSPs in close binaries (P$_{orb}<1\,$day; redbacks and black widows), searching for optical periodic signals with the \sifap\ photometer mounted on the \textit{Telescopio Nazionale Galileo} (TNG).}
   {We performed pulsation searches using the epoch-folding method with the most up-to-date (radio or gamma-ray) orbital solution for each system. These solutions were based either on previously published works or recent observations from Nançay, the Giant Metrewave Radio Telescope, the Parkes Observatory, and \fermi-LAT. 
   }
   {We update or confirm radio and gamma-ray ephemerides for five of the presented systems. No optical pulsations were detected with a significance exceeding a $3\sigma$ confidence level in any of our searches, despite achieving sensitivities close to the instrument's limit. This places correspondingly stringent upper limits on the systems' optical pulsed magnitude and conversion efficiency.
   Our upper limits are consistent with the pulsed optical luminosity of optical pulsations from the redback PSR J2339-0533.
   Compared to accreting and transitional MSPs, redback and black widow systems show much lower efficiency in converting their spin-down power into pulsed optical emission. This supports the recent suggestion that the accretion disk around the neutron star plays a role in accelerating the particles responsible for producing bright optical pulsed emission.
   }
   {}
  

   \maketitle
%

\section{Introduction}

Millisecond pulsars (MSPs) are neutron stars (NSs) rotating with a spin frequency exceeding $100$ Hz. A fraction of the MSP population exists in tight binary systems (with orbital periods $\mathrm{P_{orb}}$ on the order of hours) with low-mass ($\lesssim1\mathrm{M_\odot}$) companion stars.
These systems are formed via a recycling phase during which matter expelled by the companion star is accreted onto the NS. During this phase, the source appears as an X-ray emitter, and the accretion process is accompanied by the transfer of angular momentum, which spins the NS up to rotational frequencies that are characteristic of MSPs \citep{Alpar1982, Bhattacharya1991}.
When the accretion phase is over, the bulk of their emission comes from rotation-powered mechanisms, mainly occurring at radio or $\gamma$-ray frequencies \citep{Backer1983, Abdo2009}.
As the rotation-powered pulsar wind emitted by the NS in such close binaries is able to eject matter off its companion's surface, these pulsars are called ``spiders.'' These are divided into subclasses according to the companion star's mass. Redbacks (RBs) are defined as binary systems hosting a donor star of $M\sim0.2-0.4\,$M$_{\odot}$, while black widows (BWs) host companions with $M<0.1\,$M$_{\odot}$. A small subset of spiders has been observed to switch between the accretion disk and rotation-powered pulsar states on timescales of weeks to months. These systems have therefore been dubbed transitional MSPs (tMSPs) due to their switching between X-ray- and radio-dominated emissions \citep{Archibald2009, Papitto2013}.

Recently, optical pulsations were discovered at the same frequency as the NS spin in a variety of binary systems hosting MSPs. 
First, optical pulsations have been detected in the tMSP, PSR J1023+0038 (\citealt[see also \citealt{Papitto2019, Illiano2023}]{Ambrosino2017}), and then independently confirmed \citep{Zampieri2019, Karpov2019}. 
PSR J1023+0038 has also been discovered to show pulsations in the UV band \citep{Miraval2022, Jaodand2021}. Subsequently, optical and UV pulsations were observed from the accreting MSP SAX J1808.4-3658 \citep{Ambrosino2021, Ballocco}. More recently, evidence for optical pulsations has been reported for the RB PSR J2339–0533 \citep{Papitto2025}.
In this regard, the Silicon Fast optical Photometer (\sifap, \citealp{Ghedina2018}) was key to enabling the discovery of these new detections.

In this work, we searched for optical pulsations from a sample of eight MSPs observed with \sifap\ between 2018 and 2025. The sample consists of 
six RBs (PSR J1048+2339, PSR J1723-2837, PSR J1816+4510, PSR J2129-0429, PSR J2215+5135, and PSR J2055+1545) and two BWs (PSR J1653-0158 and PSR J1959+2048).
We used our findings to constrain the efficiency of the conversion of their spin-down power into optical pulsed emission, which is key to understanding the mechanism producing optical pulsations in rotation-powered MSPs compared to other MSP systems, such as accreting and tMSPs.

\section{Observations}\label{sec:observations}

We performed our observations with the improved version of the Silicon Fast Astronomical Photometer and Polarimeter, \sifap \citep{Meddi2012,Ambrosino2014, Ghedina2018}. \sifap\ is a non-imaging fast photometer mounted at the 3.6-m Telescopio Nazionale Galileo (TNG; \citealt{Barbieri1994}). 
It consists of two detectors covering a field of view of $7\times7$ arcsec$^2$ centered on the source and reference star (or background sky) positions, respectively. \sifap\ records the arrival times of individual photons with an 8 ns resolution.
Instrumental calibration was performed using on-site observations in white light of a series of K0V-K5V stars observed at small zenith distances. Then, these were compared with their cataloged V-magnitude values\footnote{\url{https://www.tng.iac.es/instruments/sifap2/}.}.
Observations were carried out in white light configuration (no filters applied), with a bandpass of 320-900 nm, peaking in the B band at $\sim450$ nm.

Background count rates for individual observations were estimated using at least one of two methods. The first involves simultaneous observations of an empty region of the sky about $4$ arcsec from the source. The second uses the nodding technique, which consists of tilting the source detector $\sim20$ arcsec away from the source to an empty region for $\sim100$ s at a time, more than once per observation.

\begin{table}[!t]
\caption{\sifap observation log used in this work.} \label{table:log}
\centering
\begin{tabular}{lcc}
\hline\hline 
 Source & Obs. Start Time & Exposure\\
& [MJD]&  [ks]  \\
\\
\color{black}\multirow{7}{*}{PSR J1048+2339} &58515.19481 & 6.3\\
& 58880.16773 & 8.9\\
& 60053.06180 & 5.5 \\
& 60055.01996 & 7.1 \\
& 60056.87847 & 21.6\\
& 60057.88958 & 20.4\\
& 60624.17888& 6.2\\
\\
PSR J1653-0158 & 59436.89409 & 10.9 \\
\\
\color{black}\multirow{3}{*}{PSR J1723-2837} & 
58692.90870 & 7.2 \\
& 58693.92016 & 2.7 \\
&59433.89791 & 10.8 \\
\\
\color{black}\multirow{4}{*}{PSR J1816+4510} & 59756.17187 & 2.6 \\
& 59756.90972 & 6.6 \\
& 59757.90763 & 6.2 \\
& 60836.05208 & 5.2 \\
\\
\color{black}\multirow{4}{*}{PSR J1959+2048} & 59435.93576 & 16.2\\
& 60475.04027 & 13.8 \\
& 60496.01180 & 16.0 \\
& 60496.97951 & 18.8 \\
\\
\color{black}\multirow{4}{*}{PSR J2055+1545} & 60968.81250 & 20.7 \\
& 60969.82534& 16.3\\
& 60970.82152 & 19.1 \\
& 60971.82604& 19.5\\
\\
\color{black}\multirow{6}{*}{PSR J2129-0429} & 58318.13544 & 3.3 \\
& 58318.18405 & 2.7 \\
& 58692.10939 & 4.2 \\
& 58693.01148 & 12.6 \\
& 58694.11148 & 9.6 \\
& 59437.89768& 14.2\\
\\
\color{black}\multirow{2}{*}{PSR J2215+5135}  & 59906.81111 & 14.4 \\
& 60178.19965& 2.6\\
\midrule
\hline 
\end{tabular}
\end{table} 

In most cases, individual sources have been observed several times at different epochs. This guaranties even coverage of the binary orbit, increasing exposure and maximizing the chance of detecting pulsation. A log of the \sifap\ observations analyzed in this work is reported in Table~\ref{table:log}.
Averaging over the wide range of observational conditions, we adopted a conservative air-mass value of 1.60 (see Sect. \ref{sec:timing}), based on monthly variability. The average seeing at the TNG site is $0.8$ arcsec and, in all cases, it remained below 1 arcsec during our observing runs. As a consequence, our instrumental sensitivity suffers from atmospheric flux attenuation, which degrades the limiting sensitivity by about 0.75 mag. 
More details on how this effect influences our results are given below in Sect. \ref{sec:timing}.

\begin{table*}[!t]
\caption{Orbital solutions for the set of pulsars analyzed in this work.} \label{table:ephemeris}
\centering
\begin{tabular}{lccc}
\hline\hline 
Parameter & {J1723} & {J2215} & {J1816}\\
\\
RA, $\alpha$ (J2000) & 17:23:23.1856(8)& 22:15:32.6869(2) & 18:16:35.9340(5) \\
Dec., $\delta$ (J2000) & -28:37:57.17(11)& 51:35:36.4355(3) & 45:10:33.866(6)\\
Reference epoch (MJD) & $55667.106584$& 57205.8750 & 59756\\
Spin frequency, f (Hz) & $538.870683489(3)$& 383.19759400788(5) & 313.17493429584(3)\\
Spin-down rate, $\dot{f}$ ($10^{-15}$ Hz s$^{-1}$ ) &  $2.2(1)$& 4.1416(6) & 4.2252(2)\\
Orbital frequency, $f$ ($10^{-5}$ Hz) &  $1.88062884(2)$& 6.709535(2) & 3.2070609919(7)\\
Project. semimajor axis, $a\sin i$/c (lt-s) &  $1.225822(9)$& 0.468131(6) & 0.595380(9) \\
Eccentricity, $e$ ($10^{-6}$)& --& -- & 8(2)\\
Argument of periastron, $w$ ($^{\circ}$)&  --& -- & 97(14)\\
Epoch of ascending node, T$_{ASC}$ (MJD)&  58694.518053(6) & 56383.67277(2) & 60835.883780(2)\\
Ephemeris origin & GMRT$^{*,6}$ & Fermi-LAT$^5$ & Green Bank$^1$\\
\midrule
 & {J2129}& {J1048}& {J1959}\\
\\
RA, $\alpha$ (J2000) &  21:29:45.05(8)&  10:48:43.4183(8) & 19:59:36.748(2) \\
Dec., $\delta$ (J2000)  & -04:29:06.81(8)& +23:39:53.404(2) & +20:48:14.564(4)\\
Reference epoch (MJD) & 55750 & 56700 & 56000\\
Spin frequency, f (Hz) & $131.31034864(8)$&  214.35478534113(2) & 622.1220164758(2)\\
Spin-down rate, $\dot{f}$ ($10^{-15}$ Hz s$^{-1}$ ) &  $5.652(9)$& 1.382(2) & 6.5113(6)\\
Orbital frequency, $f$ ($10^{-5}$ Hz) &  $ 1.822810115(6)$& 4.620035550962(5) & 3.02989829702(1)\\
Project. semimajor axis, $a\sin i$/c (lt s) &  $1.85219(2)$&  0.836120(3) & 0.0892247(4)\\
Eccentricity, $e$ ($10^{-5}$) & --& -- & 4.9(6)\\
Argument of periastron, $\omega$ ($^\circ$) &  --& -- & $122(7)$\\
Epoch of ascending node, T$_{ASC}$ (MJD)&  55702.111161(7) & 56637.598174(2) & 58135.6012543(3)\\
Ephemeris origin & GBT$^2$/Fermi-LAT$^5$ & Fermi-LAT$^5$ & Nançay$^{3,6}$ \\
\midrule
 & {J2055}& {J1653}\\
\\
RA, $\alpha$ (J2000) &  20:55:47.83348(4) & 16:53:38.0525(1) \\
Dec., $\delta$ (J2000)  & +15:45:21.1988(6)& -01:58:36.903(5) \\
Reference epoch (MJD) &  $61088$&  57520.922447 \\
Spin frequency, f (Hz) & $ 463.17144340(2)$& 508.21219449806(1) \\
Spin-down rate, $\dot{f}$ ($10^{-15}$ Hz s$^{-1}$ ) &  $2.399(3)$& 0.6203(2)\\
Orbital frequency, $f$ ($10^{-5}$ Hz) & 5.76612182(4)& 22.281505(2) \\
Project. semimajor axis, $a\sin i$/c &  $0.59969(2) $& 0.010704(8)\\
Eccentricity, $e$ ($10^{-5}$)& --& -- \\
Argument of periastron, $w$ ($^{\circ}$)&  --& -- \\
Epoch of ascending node, T$_{ASC}$ (MJD)& 61098.0125490(5)  & 56513.479159(6) \\
Ephemeris origin & Arecibo$^4$/Parkes$^6$ & Fermi-LAT$^{5, 6}$\\
\hline\hline 
\end{tabular}
\tablefoot{
References: $^*$Ridolfi, A. priv. comm., $^1$\citet{Fiore2023}; $^2$\citet{Bangale2024}; $^3$\citet{Blanchard2025}; $^4$\citet{Lewis2023}; $^5$all others: \citet{Smith2023}; $^6$updated in this work.}
\end{table*}

\sifap\ observations were scheduled based on the source's observability, optical luminosity, variability, distance, and spin-down power, prioritizing the targets that improve the chances of detecting optical pulsations. 
Here, we report only on \sifap\ observations of spider MSPs that accumulated sufficient exposure by mid-October 2025.
When available for our analysis, we used the distance values obtained from Gaia data release 3 (DR3; \citealt{Bailer-Jones21}). 
Otherwise, we used the distances derived from the dispersion measure of the radio counterpart, based on the \citet[YMW16 hereafter]{Yao2017} or \citet[NE2001 hereafter]{Cordes2001} models of Galactic electron density.
We also refer to the Spider Catalog \citep{Koljonen2025} for more details on the sources analyzed in this work.

\section{Timing analysis}\label{sec:timing}

\subsection{Arrival times corrections}

To search for pulsed signals during our \sifap observations, we first corrected all photon arrival times to the Solar System barycenter frame of reference using the DE405 ephemeris. We then considered the most updated orbital ephemeris for each source to correct the arrival times for the pulsar binary motion and improve sensitivity to coherent pulsating signals.
For PSR J2215+5135, J2129-0429, J1048+2339, and J1653-0158, we obtained the initial ephemerides from the publicly available third pulsars catalog (3PC) \fermi-LAT solutions \citep{Smith2023}.
The ephemerides used for the timing analysis are reported in Table~\ref{table:ephemeris}, while additional details on the search parameters and the system's properties are listed in Table \ref{table:search_par} and Sect. \ref{sec:sample}.
In two cases (PSR J1653-0158 and PSR J1816+4510), the available solution was stable enough to be propagated to our \sifap\ observing times without producing uncertainties in any of the relevant timing parameters larger than our search sensitivity.
However, other sources from our sample showed strong orbital timing noise, as is often the case for spider MSPs (see, e.g., \citealt[and references therein]{Ridolfi2016}). This is possibly due to the cyclic modulation of the orbital period due to gravitational quadrupole moment exchange between the binary components \citep{Pletsch+Clark2015, DeFalco2025}.
Our \sifap\ exposures are too short to be sensitive to orbital period modulations (on the order of $\Delta P_{orb}/P_{orb}\sim10^{-7}$; see, e.g., \citealt{Pletsch+Clark2015, Clark2021}). However, when left unmodeled, such modulations cause the pulsar's time of passage at the orbit's ascending node ($T_{ASC}$) to deviate from a pure propagation of the initial ephemeris, which assumes a constant orbital period.
Due to such variations, a given ephemeris is typically considered valid only for a specific time period.
Within the validity period of the orbital solution, typical variations in the $T_{ASC}$ value around the average solution are on the order of $\Delta T_{ASC}\sim10\,$s or greater \citep[see also, e.g.,][]{Rosenthal2025}.
If this is the case, to correct the arrival times, we cannot consider the $T_{ASC}$ parameter as fixed. Instead, our periodicity search must also account for possible variations in $T_{ASC}$. Beyond the validity range of the ephemeris, we assumed that variations in $T_{ASC}$ would be on the order of $\Delta T_{ASC}$ (that is, tens of seconds on average, though the actual value varies for each source).
The other orbital parameters (i.e., the orbital period, P$_{orb}$, and the projected semimajor axis of the pulsar orbit, $a\,\sin\,i/c$), however, do not typically show variability over the considered timescales and were therefore treated as constant in our search \citep{Rosenthal2025}.
The importance of the $T_{ASC}$ uncertainty over other orbital parameters can also be appreciated by inspecting the derivatives of the phase model for a circular orbit (see, e.g., \citealt{Leaci2015}). Additional details on the analysis chain are provided in \citet{Ambrosino2017, Ambrosino2021, Papitto2025}. The latter, especially, highlights the potential for pulsation detections from our class of targets under realistic observing conditions.

\subsection{Spin periodicity search}\label{sec:timing_search}

Based on the considerations in Sect \ref{sec:timing}, we established our $T_{ASC}$ search grid based on simulations by \citet{Caliandro2012}. They find that, to detect pulsations, the search step on $T_{ASC}$ ($dT_{ASC}$) must be on the order of 1\,s. This is demonstrated by their Eq.\,68, assuming an $\epsilon^2$ factor of $80\%$ and their normalization fitting function $S(e, W) = U_0 = 0.41$ (given the spiders' nearly perfect circular orbits, where the eccentricity is $e=0$ and the longitude of periastron $W$ is conventionally set to $0^\circ$).
The $\epsilon^2$ factor represents the power fraction that can be retained, depending on how precisely the orbit is known.
Here, we notice that the framework established by \citet{Caliandro2012} computes the $dT_{ASC}$ search step assuming a sinusoidal signal. For non-sinusoidal profiles, uncertainties on orbital parameters result in faster decoherence of the pulse shape. To account for this effect and ensure the recovery of enough signal power during the epoch-folding process in the absence of a formal quantitative adaptation for complex profiles, we conservatively reduced the search step to $dT_{ASC}/2$.

The periodicity search was performed using the epoch-folding method \citep{Leahy+83}.
Prior to the $T_{ASC}$ search, we verified that the spin period value was known with an uncertainty smaller than $\Delta\,P_F/(2m)$, where $\Delta\,P_F=P_s^2/T_{obs}$ is the Fourier period resolution, $m=32$ is the number of pulse phase bins (chosen to sample the narrow peaks in the optical, radio, and $\gamma$-ray pulse profiles from rotation-powered pulsars; see also \citealt{Papitto2025}), $P_s$ is the spin period, and $T_{obs}$ is the observational exposure.
Accordingly, no additional trials were required for the periodicity search.

Our search resulted in a set of $\chi^2$ values for each value of $T_{ASC}$ used to correct the observed time series. The $\chi^2$ values thus obtained were then divided by a factor $r=1.15$ to account for the impact of correlated noise affecting the \sifap\ detectors \citep{Papitto2025, LaPlaca2026}.
We set the detection threshold as the $\chi^2_{det}$ value with a low probability ($p<2.7\times10^{-3}$, corresponding to a 3$\sigma$ confidence level) of being attained or exceeded if the time series contained only counting noise. This accounts for the total number of trials made and assumes that the noise statistic follows a $\chi^2$ distribution with $m-1$ degrees of freedom \citep{Leahy87}.

Our analysis did not yield any signals exceeding the defined threshold. In Table \ref{table:search_par}, we provide the maximum (peak) statistics for each search, along with the number of trials.
To evaluate the upper limits at the 95\% confidence level on the signal strength, we used Eq.~(19) of \citet[derived from \citealt{Groth1975}]{Vaughan1994}. This equation accounts for the fact that, in the case of power spectra, the distribution of total power in a given frequency bin -- when both signal and noise are present -- deviates from a $\chi^2$ distribution. 
This occurs because the total observed power is the square of the sum of the signal and noise power vectors. 
In the case of epoch folding, the same relation holds due to the sum of the contributions from different phase bins (see La Placa et al., in prep.). The upper limits on the signal strength thus obtained were transformed into upper limits on the sinusoidal pulse amplitude by using the relation $A^{UL} = \sqrt{2\chi^2_{UL}/N_{\gamma}}$ \citep{Leahy87}, where $N_\gamma$ is the total number of photons recorded in the considered time series. To establish a reliable upper limit on the source pulsed amplitude, we corrected the amplitude $A^{UL}$ for the background contribution to $N_{\gamma}$ to obtain $A^{UL}_{net} = A^{UL} (1-{r_{bkg}}/{r_{tot}})^{-1}$, where $r_{tot}$ and $r_{bkg}$ are the total and background count rates, respectively.
Following the \sifap\ calibration curve\footnote{\url{https://www.tng.iac.es/instruments/sifap2/}.}, we converted the upper limits on the net pulsed rate $A^{UL}_{net}\times\,(r_{tot}-r_{bkg})$ into visual magnitudes (m$_V$). 
Next, we applied a magnitude correction to the obtained value to account for interstellar absorption and atmospheric extinction.
Interstellar absorption was estimated using HI maps \citep{HI4PI2016} to obtain the neutral hydrogen N$_H$ value in the direction of each source. We then converted the obtained N$_H$ to an extinction value in the V band, following the work by \citet{Guver+Ozel2009}.
The atmospheric extinction depends on the air mass at the telescope location, which averages to 1.60 and corresponds to a visual extinction of 0.26(1) mag. 
We report the m$_V$ values in Table \ref{table:search_par} as upper limits on the pulsed signal.
Moreover, to compare our derived magnitude values with those reported in previous studies (see, e.g., \citealt{Ambrosino2017}), we converted the m$_V$ upper limits into B-band luminosities. Our conversion assumes that the emitted source spectrum is flat (see below) and uses the distance values reported in Sect. \ref{sec:sample}.

The magnitude values reported in Table \ref{table:search_par} are solely statistical. Systematic uncertainties due to distance, extinction, and instrumental calibration are not accounted for. We, however, estimate values for each of the contributing effects. Following the \sifap\ calibration curve\footnote{https://www.tng.iac.es/instruments/sifap2/}, we estimate an instrumental calibration systematic uncertainty of about $1.5\%$ in the resulting apparent magnitude. Distance uncertainties contribute to the systematic uncertainty in the resulting magnitudes by up to about $1.5\%$, depending on the source. Systematic uncertainties on the extinction values are typically considered to affect the recovered magnitude at the $0.1\%$ level \citep{Fitzpatrick2011}.
Finally, we estimated the systematic uncertainties arising from the flat spectrum assumption, which depend on the companion's stellar type. 
Following \citet{Strader19}, we assumed a stellar surface temperature of $5600\,$K.
Blackbody emission from such a star would therefore peak in the visible band.
In this case, fluxes in the B and V bands would only differ by about $3\%$, and their magnitude difference could be as high as about $1.5\%$. In total, we estimate that systematic uncertainties may alter our results in the B-band luminosity by up to a factor of three. Nonetheless, even when accounting for systematic uncertainties, our findings remain valid.

\begin{figure}[!t]
    \centering
    \includegraphics[width=\columnwidth]{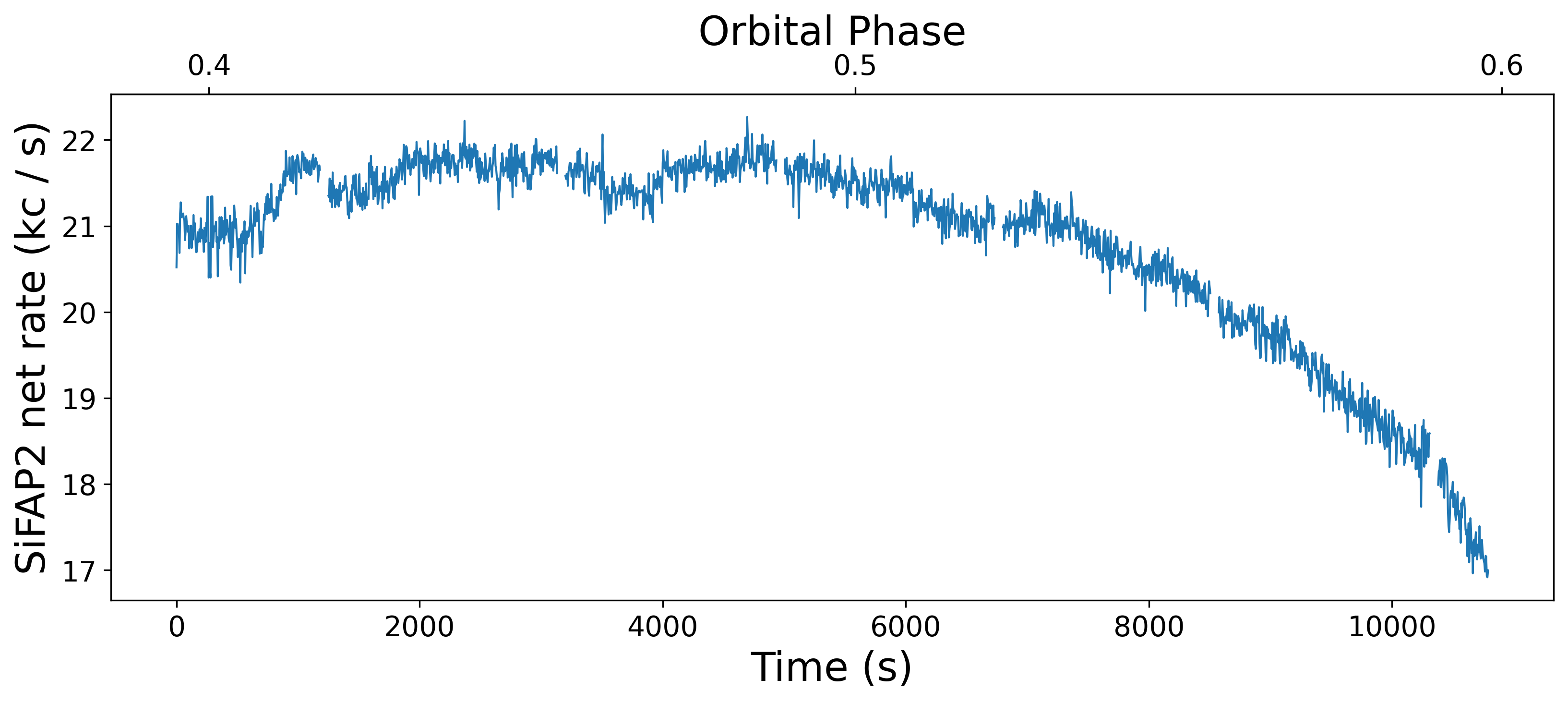}
    \caption{\sifap\ light curve from J1723, as observed in 2021. The resulting light curve is rebinned with 5 s bins, background-subtracted (estimated using the nodding technique), and corrected for dark noise. The y-axis units are in $10^3\,$c/s, while the top x-axis shows the orbital phase according to the solution in Table \ref{table:ephemeris}. The bottom x-axis indicates the time elapsed since the start of the observation (MJD $\sim59434$). The top x-axis represents the orbital phase with respect to the time of ascending node. See \citet{Turchetta2023} for a comparison of different optical light curves from RB systems.}
    \label{fig:j1723_lc}
\end{figure}

\section{The \sifap\ sample of MSPs}\label{sec:sample}

\subsection{PSR J1048+2339}\label{subsec:J1048}

PSR J1048+2339 (J1048 hereafter) is an RB system with a 4.7 ms pulsar and an orbital period of about 6 hr \citep{Cromartie2016}. It has a spin-down power of $1.2\times10^{34}\,$erg\,s$^{-1}$ and is located at a distance of 1.7 kpc \citep{Deneva2016}.
J1048 was detected by \fermi-LAT and is listed in the 3PC catalog with an orbital solution \citep{Deneva2021}. This solution includes several orbital period derivatives that are used to model the stochastic T$_{ASC}$ variability over the validity range. 
However, the period of validity does not cover our \sifap\ observation.
As an estimate of the number of trials for a search over the T$_{ASC}$, we proceeded as follows. We first propagated the value of the T$_{ASC}$ and its uncertainty to the \sifap\ observing times, assuming a constant orbital period (i.e., neglecting the orbital period derivatives present in the 3PC solution). We then estimated, within the \fermi-LAT solution validity period, the variability range of the orbital period relative to a constant value. 
This value was then added in quadrature to the propagated uncertainty. Finally, we considered a multiplying factor of 6 to account for a $3\sigma$ span on each side of the propagated T$_{ASC}$ value.

\subsection{PSR J1653-0158}

PSR J1653-0158 (J1653 hereafter) is a BW system with a 1.9 ms pulsar and a 1.2 hr orbital period \citep{Nieder2020}. The binary orbit is among the most compact systems of its type. J1653 was detected by \fermi-LAT and has a solution in the 3PC catalog, whose behavior is remarkably stable.
In fact, the \fermi-LAT solution does not require an orbital period derivative. Although the published 3PC solution is limited to a time range that does not include our \sifap\ observation times, the source orbit solution based on \fermi-LAT data remains stable and can also be adopted at later times (that is, at least up to MJD 60362; M. Kerr priv. comm.).
Following \citet{Nieder2020}, we employed a distance of 0.84(4) kpc, consistent with Gaia DR3 \citep{Koljonen2023}.
Also, we adopted a spin-down power value of $1.2\times10^{34}\,$erg/s \citep{Romani2014}.

\begin{table*}[!t]
\caption{Parameters used for the periodicity search with \sifap\ on sources analyzed in this work and our main search results.} \label{table:search_par}
\centering
\begin{tabular}{lccccccccc}
\hline\hline 
 Source & T$_{ASC}$ & P$_{spin}$ & Search range$^*$ & Search step & \multicolumn{2}{c}{Peak statistics$^{\ddagger}$} & Trials & m$_V^\dagger$ \\
        &   [MJD]   &  [ms]       &  [seconds]   & [seconds]  & $\chi^2$ & $\chi^2/r$ & & [mag]\\ 
        \\
\color{black}\multirow{4}{*}{J1048} 
& 58514.98873(19) & 4.665162940467(4) & 71 & 2.5 & 50.6 & 44.0 & 57 & \\
& 58879.99514(24) & 4.665162941415(5) & 78 & 2.5 & 52.4 & 45.6 & 63 & \\ 
& 60052.92583(35) & 4.665162944462(8) & 105 & 2.5& 51.2 & 44.5 & 83 & \\
& 60624.10950(42) & 4.665162945946(9) & 119 & 2.5& 53.6 & 46.6 & 95 & 25.6\\
\\
\color{black}\multirow{1}{*}{J1653} & 59436.8781454(7) & 1.96768202539829(4) & 66 & 17 & 43.3 & 37.6 & 5 & 25.5\\
\\
\multirow{2}{*}{J1723} & 58694.518053(6) & 1.85573279571521(1)& -- & 1.6 & 33.6& 29.2& 1 & 23.0\\
& 59433.65715(2) & 1.855732798898(1)& 50 & 1.6 & 60.5 & 52.6 & 63 & \\
\\
\color{black}\multirow{2}{*}{J1816} & 59756.090483(1) & 3.1931035676546(3) & -- & 3.4 & 53.0 & 46.1 & 1 & 25.4\\
& 60836.052083(2) & 3.1931035676544(3) & -- & 3.4 & 24.5 & 21.3 & 1 & \\
\\
\color{black}\multirow{3}{*}{J1959} & 59435.81609(65) &                      1.6074017260663(1) & 60 & 12 & 54.6 & 47.5 & 5 & 24.5 \\
& 60475.03(12) & 1.6074017275759(2)  & 2592 & 12 & 66.3 & 57.6 & 5185 & \\
& 60496.04(13) & 1.6074017276064(2) & 2808 & 12 & 72.6 & 63.1 & 5617 & \\
\\
\multirow{1}{*}{J2055} & 60968.745358(64) & 2.159027751340898(1)& 16 & 1.3 & 54.8 & 47.6 & 27 & 26.2\\
\\
\color{black}\multirow{3}{*}{J2129} & $58317.97895(8)$ & 7.61393754314(1) & 9 & 4.5 & 42.4 & 36.9 & 5 & \\
& $58692.12809(9)$ & 7.61393755373(2)& 9 & 4.5 & 53.3 & 46.3 & 5 & 25.7\\
& $59437.8854(1)$&   7.61393757485(4)& 148 & 4.5 & 52.6& 45.7& 67 & \\ 
\\
\color{black}\multirow{2}{*}{J2215} & 59906.6715(4) &                         2.609619733909(1) & 6 & 1.7 & 67.4 & 58.6 & 5 & 24.3 \\
& 60178.1894(4) & 2.609619734571(2) & 6 & 1.7 & 59.5 & 51.7 & 5 & \\
\\
\midrule
\hline 
\end{tabular}
\tablefoot{
$^{*}$Search range around the T$_{ASC}$ $\pm$ indicated values (not listed for single trial searches). $^\ddagger$The $\chi^2$ values corresponding to the peak of the periodicity search and those corrected for the detector's correlated noise. The values used for the conversion into magnitude upper limits (not shown here) also account for the $\chi^2_{noise}$ and the interaction between signal and noise powers (see Sect. \ref{sec:timing_search}). 
$^{\dagger}$The visual V-band magnitude upper limit on the pulsed signal, reported only for the corresponding (trial-corrected) highest statistics.}
\end{table*}

\subsection{PSR J1723-2837}

PSR J1723-2837 (J1723 hereafter) is an RB system exhibiting radio eclipses \citep{Crawford2013}.
There is no significant $\gamma$-ray detection for this source, and a \fermi-LAT solution for it is not publicly available.
The NS spin period is 1.86 ms \citep{Faulkner2004}, and the orbital period is about 15 hr.
Its Gaia DR3 distance is 0.9 kpc \citep{Koljonen2023}, while its spin-down power is $4.6\times10^{34}\,$erg\,s$^{-1}$ \citep{Crawford2013}.
It was recently observed with the \ixpe\ observatory as part of an extended observational campaign, although no significant X-ray polarization has been detected \citep{Negro2026,Sullivan2026}.
The most recent timing solution for this source was published by \citet{Crawford2013}. However, this solution does not cover our 2019 and 2022 \sifap\ observations of the source.
For this reason, Giant Metrewave Radio Telescope (GMRT) observations were performed (Ridolfi et al., priv. comm.) to obtain a radio timing solution 
that covers our \sifap\ 2019 observations (see Tables~\ref{table:ephemeris} and \ref{table:search_par}).
We therefore employed this solution to search for pulsations in our 2019 observations.
Since a single radio observation does not allow for a reliable fit of the arrival times, the radio data were simply folded using the \citet{Crawford2013} solution, adjusting T$_{ASC}$ in small steps until the highest S/N ratio was reached. This method does not provide uncertainties for the fitted parameters. Instead, we used those reported by \citet{Crawford2013}.
Moreover, we propagated the T$_{ASC}$ obtained through the GMRT observations to the day of our 2021 \sifap\ observation (not covered by radio observations), assuming an uncertainty similar to the orbital phase variability observed during the validity interval of the \citet{Crawford2013} solution. This corresponds to a conservative range of 50 s around the propagated T$_{ASC}$.
A light curve of the J1723 \sifap\ observation is shown in Fig. \ref{fig:j1723_lc}.

\subsection{PSR J1816+4510}\label{subsec:J1816}

\begin{figure}[!t]
    \centering
    \includegraphics[width=\columnwidth]{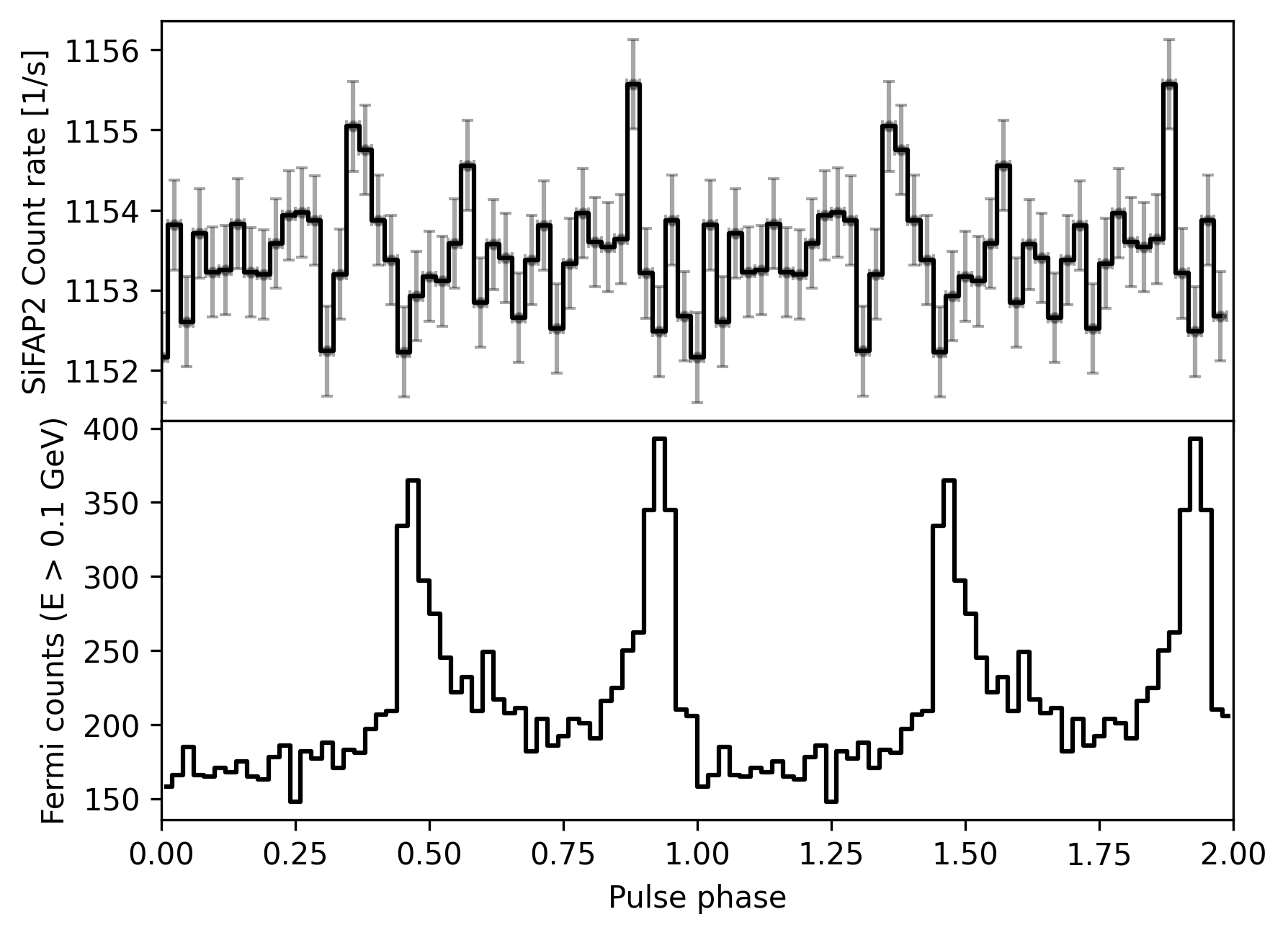}
    \caption{PSR J1816 pulse profiles, as observed by \sifap\ (top panel) and Fermi-LAT (bottom panel). The reference epoch time for both profiles is MJD 59756. The \sifap\ count rates are not background-corrected. Although our search for pulsation does not return a significant signal (see Sect. \ref{subsec:J1816} and Table \ref{table:search_par}), the two profiles appear similar. The difference in binning between the profiles is due to the differing available statistics.}
    \label{fig:J1816}
\end{figure}

PSR J1816+4510 (J1816 hereafter) is an RB system with a 3.2 ms pulsar and an 8.7 hr long orbital period \citep{Ray2012, Kaplan2012}. It was also observed and characterized by the Green Bank Northern Celestial Cap Pulsar Survey \citep{Stovall2014}. The system was also recently observed with the Five-hundred-meter Aperture Spherical radio Telescope (FAST) to study its radio eclipse behavior \citep{Shang2024}.
The adopted source distance is 4.36 kpc \citep{Yao2017, Shang2024}, and the spin-down luminosity is $5\times10^{34}\,$erg/s \citep{Fiore2023}.
Similar to J1653, J1816 has a publicly available \fermi-LAT 3PC timing solution that is remarkably steady and can be extended up to at least MJD 60884 (D. Smith, M. Kerr, priv. comm.).
Given this solution, we employed \texttt{PRESTO}\footnote{\url{https://github.com/scottransom/presto}.} \citep{Ransom2011} to fit \fermi-LAT data, obtaining a precise timing solution that covers our \sifap\ observations. This allowed us to perform single trial searches for pulsations using the best-fit T$_{ASC}$ nearest to our \sifap\ observations.
Our periodicity search from J1816 yields a $2.1\sigma\,$confidence level signal which, although of low significance, is relevant among those obtained in this work.
We therefore produced a \sifap\ pulse profile of the 2022 observations (see Fig \ref{fig:J1816}). We compared this to the unweighted \fermi-LAT profile above 0.1 GeV obtained with a $0.8^\circ$ radius of interest. Although $\gamma$-ray pulse profiles can differ from those obtained in different energy bands (see, e.g., \citealt{Abdo2010b}), their shape similarity, peak separation, and approximate phase alignment are intriguing.

\subsection{PSR J1959+2048}

PSR J1959+2048 (also dubbed PSR B1957+20; J1959 hereafter) is a BW system with a spin period of 1.6 ms and an orbital period of 9.2 hours \citep{Fruchter1988}. 
For this source, we adopted a spin-down luminosity value of $1.6\times10^{35}\,$erg/s \citep{Koljonen2025} and a distance value of $1.73\,$kpc \citep{Yao2017}. 
We employed the ephemeris obtained by the Nançay telescope around our \sifap\, observing times (see \citealt{Blanchard2025}).
These were fitted with the 
ELL1 model \citep{Lange01}, which utilizes the Laplace-Lagrange parameters while also accounting for the possible small eccentricity of the orbit.
Given the large uncertainties derived from the propagation of the T$_{ASC}$ to the epochs of the 2024 \sifap\ observations, a large number of trials are required for the periodicity search (see Table \ref{table:search_par}).
These span a $3\sigma$ interval around the central value.

\subsection{PSR J2055+1545}

PSR J2055+1545 (J2055 hereafter) is an RB system, discovered by \citet{Lewis2023} in the Arecibo 327 MHz Drift-Scan Pulsar Survey. It is the most recently discovered source among those analyzed in this work.
Although a coincident \fermi-LAT counterpart was detected for this source, no $\gamma$-ray pulsations were found. J2055 consists of a NS with a 2.2 ms spin period in a circular 4.8 hr orbit \citep{Lewis2023}. Following \citet{Koljonen2025} we adopted the YMW16 distance of 3.7 kpc, along with its spin-down power of $4.0\times10^{34}\,$erg/s \citet{Lewis2023}. 
The system shows radio eclipses as observed by the Arecibo survey \citet{Lewis2023}.
For this source, we obtained radio observations using the Murriyang radio telescope in Parkes (NSW, Australia) at the request of Green Time under project PX090. Its ultra-wideband low receiver (704-4032 MHz; \citealt{2020PASA...37...12H})  was used in search mode, with 1-megahertz-wide channels coherently de-dispersed at the pulsar's dispersion measure. 
Three epochs were scheduled for February 7, 26, and March 30, 2026, at times when the source was predicted to be outside the radio eclipse. 
Each observation was folded starting from the ephemeris of \citealt{2023ApJ...956..132L}, with a T0 range covering $\pm$ 0.01 in orbital phase, using the software \texttt{spider-twister}\footnote{\url{https://alex88ridolfi.altervista.org/pagine/pulsar_software_SPIDER_TWISTER.html}}, which is based on \texttt{PRESTO} \citep{Ransom2011}. For each observation, the highest signal-to-noise folded data and the related T0 were used as starting points for the subsequent timing analysis. The best local set of ephemeris was obtained using \texttt{TEMPO2} \citep{Hobbs2006} to fit the orbital parameters. Results are shown in Table \ref{table:ephemeris}.

\subsection{PSR J2129-0429}

PSR J2129-0429 (J2129 hereafter) is an RB system with a 7.6 ms spin period and a 14.4 hr orbital period around a relatively bright (m$_R\sim16.6$ mag) optical counterpart \citep{Hessels2011}. The source was previously identified as a $\gamma$-ray source with Fermi-LAT since the first catalog \citep{Abdo2010b}.
The spin-down power is $3.9\times10^{34}\,$erg/s \citep{Roberts2014, AlNoori2018A}.
A distance value of $1.9_{-0.2}^{+0.3}\,$kpc was assumed from Gaia DR3.
Relatively updated radio orbital ephemerides obtained with the Green Bank Telescope (GBT) by \citet{Bangale2024} were employed to propagate the T$_{ASC}$ value to our \sifap\ observation periods, as the validity range of their solution covers our \sifap\ observations at least in part. Moreover, the solution from \citet{Bangale2024} coincides with that published by \fermi 3PC regarding the spin frequency and its derivative.  
Beyond the validity range, we extrapolated the radio timing solution to the epoch of the \sifap\ observations and then expanded our search to cover the variability span of the \fermi-LAT timing solution (that is, 150 s around the T$_{ASC}$ value obtained through propagation, assuming a constant orbital period).

\subsection{PSR J2215+5135}

PSR J2215+5135 (J2215 hereafter) is a RB system with a spin period of 2.6 ms and an orbital period of 4.1 hr \citep{Hessels2011, Bangale2024}.
We adopted a distance value of 3.33 kpc for this source \citep{Sullivan2024} and a spin-down power of  $5.3\times10^{34}\,$erg/s \citep{Breton2013}.
To calculate the T$_{ASC}$ value around our \sifap\ observations, we propagated the public \fermi-LAT orbital solution -- similarly to J1048 (see Sect. \ref{subsec:J1048}) -- assuming a constant orbital period.
The solution, however, includes a series of ORBIFUNC parameters representing a time-domain interpolating function to model the orbital phase variations over time.
Using these, the T$_{ASC}$ oscillates over a $\pm6\,$s range with respect to the T$_{ASC}$ value obtained from the initial constant orbital period ephemeris (see, e.g., \citealt{Clark2021}).
Similar to J1048, we thus added this value in quadrature to the propagated uncertainty to obtain a more conservative search range around the central T$_{ASC}$ value.

   \begin{figure*}
   \sidecaption
   \includegraphics[width=12cm]{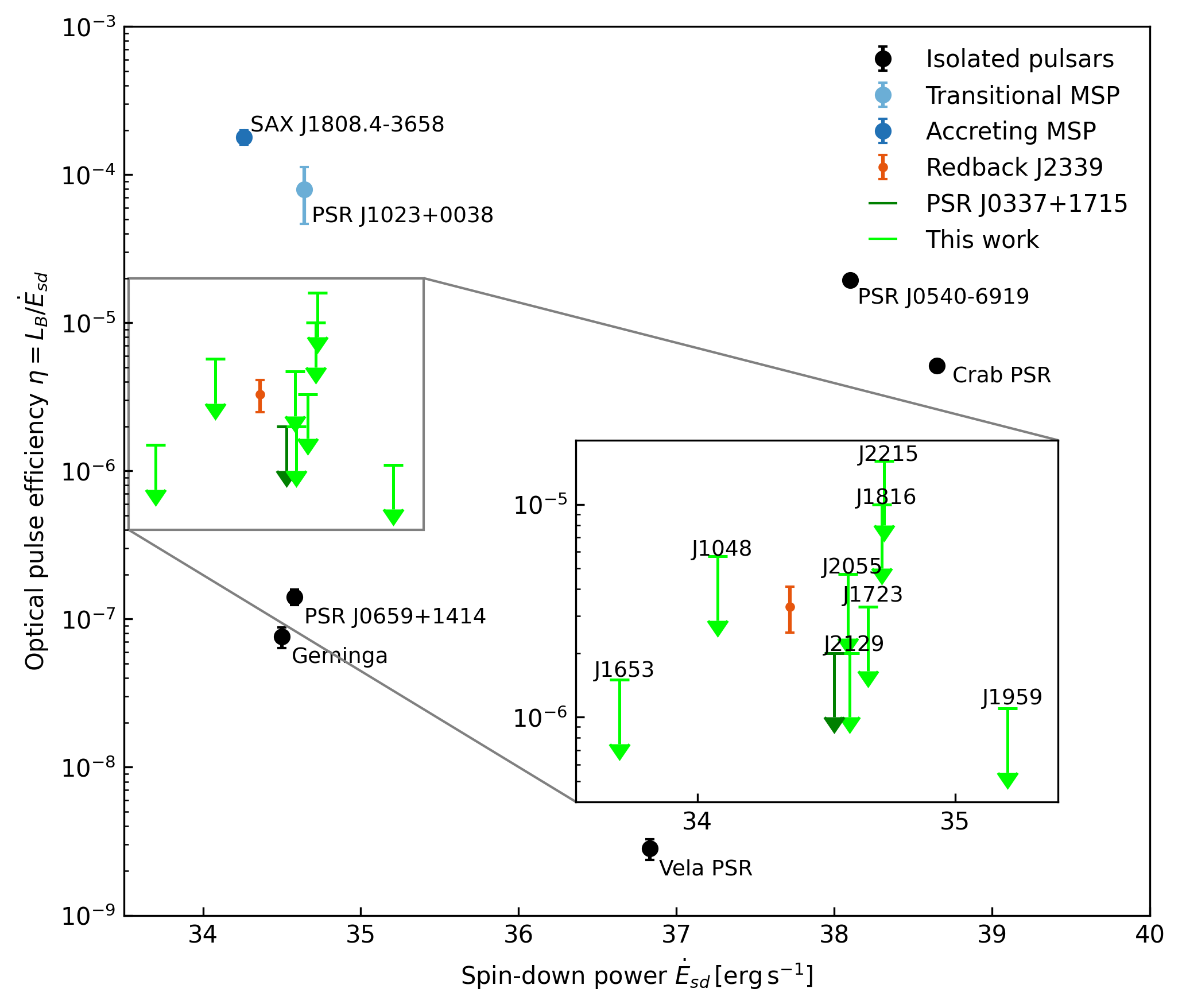}
   \caption{Optical pulse efficiency vs. spin-down power for a sample of sources. The optical pulse efficiency is defined as the ratio of the pulsed B-band luminosity ($\lambda_{eff}$ = 433 nm, W$_{eff}$ = 90 nm) to the spin-down power. The sample of MSPs analyzed in this work is represented by the downward light-green arrows, which indicate the $95\%$ confidence level upper limits on the optical pulse efficiency. See \citet{Papitto2025} and references therein for a comparison with previously detected optical pulsars (indicated in the legend, marked with data points). Compared to \citealt{Ambrosino2021}, we corrected the data point for PSR J0540-6919 using the spin-down power reported by \citealt{Ferdman2015} and the data point for SAX J1808.4-3658 using an updated de-reddening correction for Las Cumbres Observatory B-band magnitude. The inset shows a zoom-in on the sample and labels only the sources analyzed in this work.}
    \label{fig:optical-efficiency}%
    \end{figure*}

\section{Discussion}

Optical pulsations have already been detected in a few systems previously identified as radio pulsators, including five young isolated neutron stars (INSs; \citealt{Cocke1969, Mignani2011}), as well as in MSPs (whether accreting, transitional, or RB; see \citealt{Ambrosino2017, Ambrosino2021, Papitto2025}).
The optical emission from INSs is commonly interpreted as incoherent synchrotron radiation produced by relativistic electron-positron pairs accelerated in the NS magnetosphere \citep{Pacini1983, Shearer2001, Mignani2011}.
Its efficiency can be defined as the conversion efficiency of spin-down luminosity into optical luminosity in a given band (e.g., B band), expressed as $\eta=L_{opt}/\dot{E}_{sd}$. For systems such as the Crab, this value is on the order of $10^{-5}$, while for other INSs it is even lower, on the order of $10^{-8}-10^{-7}$ (see Fig. \ref{fig:optical-efficiency}). On the other hand, the optical pulse efficiency for transitional and accreting MSPs is on the order of $10^{-5}-10^{-4}$, with an optical pulsed amplitude on the order of $1\%$ or less.

In Fig. \ref{fig:optical-efficiency}, we report our optical observational campaign results over the sample of MSPs described in Sect.\,\ref{sec:sample}. Our results are reported in terms of the conversion efficiency of the spin-down power into pulsed optical luminosity. 
As we do not find any signals above the detection threshold, we are only able to establish upper limits on the optical pulse efficiency of the analyzed systems.
We nonetheless highlight these as the most stringent constraints on this type of signal from the considered sources.
We also note that the upper limits we obtained on the efficiency values are about the same as those for PSR J0337+1715, a MSP (2.7 ms spin period) in a triple hierarchical system with two white dwarfs observed with a different detector, namely ARCONS \citep[see Fig. \ref{fig:optical-efficiency}]{Strader2013}.
Moreover, they are in agreement with the results for the RB system, PSR J2339-0533, where a $3.5\sigma$ pulsation detection was found with a pulsed optical luminosity of m$_B=27.0$ mag \citep[and Fig. \ref{fig:optical-efficiency}]{Papitto2025}.
Our results support the hypothesis that the optical pulse efficiency in spiders is $\lesssim10^{-5}$, i.e., up to a couple of orders of magnitude lower than that observed for other types of MSPs in binary systems, namely transient and accreting systems (see PSR J1023+0038 and SAX J1808.4-3568 in Fig. \ref{fig:optical-efficiency}).
This is in agreement with recent models \citep{Papitto2019, Veledina2019} that highlight the role of an accretion disk in producing bright optical pulsations as synchrotron emission at the intrabinary shock, where the pulsar wind interacts with the disk \citep{Richard-Romei2026}. 
The upper limits obtained in the present work indicate that if optical pulsations are present in spider MSPs, they are generally fainter than in the currently known accreting and tMSP cases. This is consistent with, but does not unambiguously prove, a role for the accretion flow or disk in enhancing optical pulsed emission.
On the other hand, similar to INSs, spider systems have no accretion flow influencing the magnetosphere, and the mechanism producing optical pulsations is expected to be the same \citep[and references therein]{Papitto2025}.
In fact, as discussed by \citet{Papitto2025}, despite a gap in spin-down power of four orders of magnitude between isolated pulsars such as the Crab and its twin in the Large Magellanic Cloud, PSR J0540-6919, and the RB system, PSR J2339-0533, their optical pulsed efficiency ($\eta$) is similar.
On the other hand, our non-detection of pulsations in spider systems could, in principle, be due to other factors besides the intrinsically low efficiency $\eta$. Among these, we mention instrument sensitivity, the geometry of the system (which may be such that the beam of optical emission does not cross our line of sight), and the uncertainty in the ephemeris. 
Moreover, we notice that pulsations from spider systems are often transient in the radio band due to scintillation and absorption by plasma ejected by the pulsar wind. This enshrouds the system, thus making an intrinsically weak signal even harder to detect \citep{MiravalZanon2021, Ridolfi2016, Thongmeearkom2024}. The non-detection of optical pulsations in our sample can therefore also be due to their transient nature. Finally, we mention that the beaming in different energy bands can also play a role in their detectability. Remarkable cases in this regard are the \fermi-LAT $\gamma$-ray pulsars, PSR J1035-6720 and PSR J1744-7619. In fact, faint radio pulsations were detected for the former only after the discovery of its $\gamma$-ray pulsations, while no radio pulsations have been detected so far for the latter \citep{Clark2021}.

\section{Conclusions}

We analyzed a sample of eight spider MSPs observed with \sifap\ in search of optical pulsed signals. We summarize our main findings as follows:

   \begin{enumerate}
      \item We find no significant pulsations in the optical band from the analyzed sample of sources. We obtained stringent ($95\%$ confidence level) upper limits on the optical pulsed efficiency. These are compatible with the efficiency value obtained for PSR J2339-0533 -- the only other RB system for which evidence of optical pulsations has been reported with an efficiency $\eta=3.3\times 10^{-6}$.
      \item Our upper limits are a factor of 50-100 below the optical pulse efficiency measured for the transitional and accreting MSP systems, PSR J1023+0038 and SAX J1808.4-3658, respectively. These findings support the hypothesis that an accretion disk around the NS increases the efficiency of producing optical pulsations, while the optical pulse efficiency of spider systems may have the same origin as the optical pulsation from  isolated radio pulsars. 
      \item Orbital timing solutions for three systems, namely J1723, J1959, and J2055, were updated using recent radio observations. Moreover, for two other systems, namely J1653 and J1816, we verified that their known orbits derived from \fermi-LAT $\gamma$-ray data are remarkably stable and remain valid as of the time of writing. Here, we reported their most recent orbital parameters.
   \end{enumerate}

Our results on the optical pulse efficiency are also summarized in Fig. \ref{fig:optical-efficiency}. They show that an improvement by a factor of two in pulsed-flux sensitivity lowers the efficiency limits by about $50\%$, which is sufficient to test whether optical pulse efficiency values similar to those measured for PSR J2339-0533 are typical of spider MSPs. However, sensitivity alone may not secure detection, as it can be affected by geometric, ephemeris, and intermittency limitations.

\section*{Data availability}

Data used in this work is available upon request. 

\begin{acknowledgements}
This work was supported by INAF (Research Large Grant FANS and GO Grant PULSE-X, PI: Papitto; ``Chasing the optical counterpart in FRBs and Magnetars with SiFAP2@TNG and SiFAP-SoFT@Cassini'', PI: Ambrosino), the Italian Ministry of University and Research (PRIN MUR 2020, Grant 2020BRP57Z, GEMS, PI: Astone), and Fondazione Cariplo/Cassa Depositi e Prestiti (Grant 2023-2560, PI: Papitto). GI is supported by a Juan de la Cierva fellowship (JDC2024-053550-I). We also thank \"O. F. \c{C}oban, D. Smith and M. Kerr for useful discussions.
Murriyang, CSIRO’s Parkes radio telescope, is part of the Australia Telescope National Facility (\url{https://ror.org/05qajvd42}) which is funded by the Australian Government for operation as a National Facility managed by CSIRO. We acknowledge the Wiradjuri people as the Traditional Owners of the Observatory site.
The Nançay Radio Observatory is operated by the Paris Observatory, associated with the French Centre National de la Recherche Scientifique (CNRS) and Université d’Orléans. It is partially supported by the Region Centre Val de Loire in France. DFT has been supported by PID2024-155316NB-I00 funded by MICIU /AEI /10.13039/501100011033 / FEDER, UE and CSIC PIE 202350E189.
\end{acknowledgements}

%
%

\bibliographystyle{yahapj}
\bibliography{references}

\end{document}